\documentclass[aps,pre,twocolumn,showpacs,floatfix,superscriptaddress]{revtex4}
\usepackage{graphicx}

\newcommand {\be} {\begin{equation}} 
\newcommand {\ee} {\end{equation}} 
\newcommand {\Be}{\begin{eqnarray*}}
\newcommand {\Ee} {\end{eqnarray*}}
\newcommand {\bey} {\begin{eqnarray}} 
\newcommand {\eey} {\end{eqnarray}} 

\begin{document}

\title{Studies of thermal conductivity in FPU-like lattices }

\author{Stefano Lepri}
\email{lepri@inoa.it}
\affiliation{Istituto Nazionale di Ottica Applicata, largo E. Fermi 6
I-50125 Firenze, Italy}
\affiliation{Istituto dei Sistemi Complessi, Consiglio Nazionale delle
Ricerche, Sez. Territoriale di Firenze, largo E. Fermi 6
I-50125 Firenze, Italy}
\affiliation{Istituto Nazionale per la Fisica della Materia-UdR
Firenze, via G. Sansone 1 I-50019 Sesto Fiorentino, Italy}

\author{Roberto Livi}
\affiliation{Dipartimento di Fisica, via G. Sansone 1 I-50019, Sesto
Fiorentino, Italy }
\affiliation{Istituto Nazionale per la Fisica della Materia-UdR
Firenze, via G. Sansone 1 I-50019 Sesto Fiorentino, Italy}

\author{Antonio Politi}
\affiliation{Istituto Nazionale di Ottica Applicata, largo E. Fermi 6
I-50125 Firenze, Italy}
\affiliation{Istituto dei Sistemi Complessi, Consiglio Nazionale delle
Ricerche, Sez. Territoriale di Firenze, largo E. Fermi 6
I-50125 Firenze, Italy}
\affiliation{Istituto Nazionale per la Fisica della Materia-UdR
Firenze, via G. Sansone 1 I-50019 Sesto Fiorentino, Italy}

\date{\today}

\begin{abstract}
The pioneering computer simulations of the energy relaxation mechanisms 
performed by Fermi, Pasta and Ulam can be considered as the first attempt
of understanding energy relaxation and thus heat conduction in lattices of
nonlinear oscillators. In this paper we describe the most recent achievements
about the divergence of heat conductivity with the system size in 1d and 2d
FPU-like lattices. The anomalous behavior is particularly evident at low
energies, where it is enhanced by the quasi-harmonic character of the lattice
dynamics. Remakably, anomalies persist also in the strongly chaotic region
where long--time tails develop in the current autocorrelation function. 
A modal analysis of the 1d case is also presented in order to gain
further insight about the role played by boundary conditions.
\end{abstract}

\pacs{63.10.+a  05.60.-k   44.10.+i}

\maketitle  

\textbf{ Almost one century ago Pierre Debjie conjectured that nonlinearities
should be  considered as a basic ingredient for explaining relaxation and
transport mechanisms exhibited by real solids. In fact, the simplest model -- a
lattice of harmonic oscillators -- yields quite an unphysical scenario: any
initial excitation does not evolve towards an equilibrium state, but frequently
returns close the initial state, while transport is purely ballistic since any
Fourier component transfers energy unaltered through the lattice at the sound
velocity. In the thirties the introduction  of nonlinear terms in a
perturbative quantum--mechanical description, allowed  Rudolph Peierls to
obtain a successful explanation of the thermodynamics of  solids at very
low--temperature. Only twenty years later Enrico Fermi, John Pasta and Stanley
Ulam  tackled the problem of studying the relaxation and transport properties
of a lattice of nonlinear classical oscillators. As discussed all over this
issue, their numerical studies have provided inspiration for an astonishingly
large amount of investigations. In this contribution we aim at surveying the
progresses and the still open problems concerning heat conduction in the FPU
model. In particular, we discuss the divergence of heat conductivity with the
system size in the 1d and 2d versions of the model.  }

\section{Introduction}

At the beginning of the 50's one of the first digital computers, MANIAC 1, was
available at Los Alamos National Laboratories in the US. It had been designed
by the mathematician J. von Neumann  for supporting investigations in several
research fields,  where difficult mathematical problems could not be tackled by
rigorous proofs~\cite{bomb}. Very soon Enrico Fermi realized the great
potential of this revolutionary computational tool for approaching also some
basic physical questions that had remained open for decades. In particular,
MANIAC 1 appeared as a suitable  instrument for analyzing the many aspects of
nonlinear problems that could not allow for standard perturbative
methods. In collaboration with the mathematician S. Ulam and the physicist J.
Pasta, Fermi proposed to integrate by MANIAC 1 the dynamical equations of the
simplest model of a crystal: a chain of classical oscillators coupled by 
nonlinear forces described by the Hamiltonian 
\begin{equation} 
H = \sum_{i=1}^N
\left[ \frac{p_i^2}{2m} + V(q_{i+1} - q_i)  \right]
\label{FPU} 
\end{equation} 
where the integer index $i$ labels the oscillators, whose displacements
with respect to equilibrium positions and momenta are $q_i$ and $p_i$, 
respectively. For the sake of simplicity Fermi, Pasta and Ulam considered 
the interatomic potentials
\begin{equation}
V(z) = \frac12 m\omega^2 z^2 + {\alpha \over 3} z^3 
\end{equation} 
and
\begin{equation}
V(z) = \frac12 m\omega^2 z^2 + {\beta \over 4} z^4 
\end{equation} 
which have been therafter termed ``$\alpha$'' and ``$\beta$'' models
respectively. The corresponding equations of motion were implemented
by M. Tsingou into a program containing an integration algorithm that 
MANIAC 1 could efficiently compute. 

As discussed elsewhere in this Focus Issue, the FPU numerical experiment was
intended to test how equilibrium (equipartition) is approached by an isolated
set of nonlinearly coupled oscillators. On the other hand, the connection of
this problem to the one of transmission of vibrational energy must not have
been escaped to Fermi's physical intuition. Already in 1914, the dutch
physicist  P.~Debye had argued that nonlinearity in the interatomic forces is
necessary for the finiteness of thermal conductivity of insulating crystals,
i.e. for the Fourier's law $\vec{J}=-\kappa\nabla T$ to hold. Since, despite
the many simplifications, this basic ingredient is included in the FPU model,
one could hope to describe this important physical effect in a concrete case.

Furthermore, the measurement of  the time scale for approaching the equilibrium
state, i.e. the``relaxation time" of plane-wave excitations, would have
provided an indirect determination of thermal conductivity. Indeed, the most
elementary picture of heat conductivity is based on the analogy with  kinetic
theory of gases where $\kappa=Cv_s \ell/3$, $C$ being the heat  capacity, $v_s$
the sound velocity and $\ell$ the mean free path. In a lattice, heat carriers
are phonons (classically the normal modes), and it is  
necessary to take into account that the latter have different group 
velocities, $v_{\bf k}={\partial \omega /\partial {\bf k}}$, depending on their
wavenumber ${\bf k}$. Accordingly, the above expression for $\kappa$ generalizes to
\begin{equation}
\kappa \; \propto \; \sum_{\bf k} \, C_{\bf k}v^2_{\bf k} \tau_{\bf k} 
\quad ,
\label{debye}
\end{equation} 
where we have introduced the relaxation time 
$\tau_{\bf k} = \ell_{\bf k}/v_{\bf k}$ that can be determined by
phenomenologically including all possible scattering mechanisms (anharmonicity, 
impurities, boundary effects, electrons etc.) that must be determined in some 
independent way. 

In their numerical experiment \cite{FPU65} FPU observed that, at variance with the original
intuition, the energy initially fed in one of the low, i.e. long--wavelength,
oscillatory mode did not flow to the higher modes, but was exchanged only among
a small number of low modes, before  flowing back almost exactly to the initial
state, yielding a recurrent behavior. Despite nonlinearities were at work,
neither a tendency towards thermalization, nor a mixing rate of the energy
could be identified. The dynamics exhibited regular features very close to
those of an integrable system. Thus, in the spirit of the Debye argument, this 
infinite relaxation time would imply an infinite conductivity also in presence
of nonlinear forces!  

This indirect consequence of Fermi, Pasta and Ulam's work received an 
intuitively appealing confirmation with the discovery of Zabusky and Kruskal's
soliton. Generally speaking, whenever the equilibrium dynamics of a lattice can
be decomposed into that of  independent ``modes'', the system is expected to
behave as an ideal conductor. Besides the trivial example of the  harmonic
crystal, this applies also to the broader case of integrable nonlinear models
characterized by the presence of ``mathematical solitons" originating the
balance of dispersion and nonlinearity. The idea that solitons may play a role
in heat conduction dates back to Toda \cite{T79} and has been invoked to
explain the anomalous behavior of the FPU model as a consequence of ballistic
transport due to solitons of the modified Korteweg--deVries equation (see
e.g.~\cite{C86}). Thereby, the existence of stable nonlinear excitations in
integrable systems is expected to lead to ballistic rather than to diffusive
transport. As pointed  out in Ref.~\cite{T79}, solitons travel freely, no
temperature gradient can be maintained and the conductivity is thus infinite.

\section{The nonequilibrium FPU model: early results}

The original FPU simulation probed the nonequilibrium transient dynamics under
the only effect of internal forces. When dealing with systems that can exchange
energy with external reservoirs, one wishes to introduce the effect of
external temperature gradients by means of nonequilibrium simulations too. In
the spirit of linear response, the two concepts should be to some extent
related, since the relaxation of spontaneous fluctuations rules the system
response.  

In the present context, the natural way to proceed consists in putting the
system in contact with two heat reservoirs operating at different temperatures
$T_+$ and $T_-$. Several methods have been proposed based on both deterministic
and stochastic algorithms\cite{LLP02}. Regardless of the actual thermostatting
scheme, after a transient,
an off-equilibrium stationary state sets in, with a net heat current flowing
through the lattice. The thermal conductivity of the chain $\kappa$ is then
estimated as the ratio between the time--averaged flux $\overline J$  and the
overall temperature gradient $(T_+-T_-)/L$. Notice that, by this latter choice,
$\kappa$ amounts to an effective transport coefficient including both boundary
and bulk scattering mechanisms. The average $\overline J$ can be estimated in
several equivalent ways, depending on the employed thermostatting scheme. 
One possibility is to directly measure the energy exchanges with the two 
baths. A more general definition (thermostat-independent) consists in averaging
\begin{equation}
J \;=\;  {a\over 2}  \sum_n ({\dot q}_{n+1} + {\dot q}_n) \, F_n 
\label{jsolid}
\quad ,
\end{equation}
that is a suitable microscopic expression appropriate in the context of
lattices with nearest--neighbour couplings \cite{note}. Here,
$F_n=- V'(q_{n+1} - q_n)$ is a shorthand notation for the force excerted by
the $n$--th on the $n+1$--th oscillator, while $a$ is the lattice spacing.

Historically, this approach has been followed some years after the
implications of the original FPU experiment were appreciated by the nonlinear
physics community. The first papers on the FPU model under steady
nonequilibrium conditions date back to the pioneering studies of Payton, Rich
and Visscher \cite{PRV67} and Jackson, Pasta and Waters \cite{JPW68}. In both
cases, the Authors considered cubic plus quartic potential terms resulting
from the expansion of the Lennard-Jones potential. To investigate the effect of
impurities in the crystal, either a disordered binary mixture of masses
\cite{PRV67} or random nonlinear coupling constants \cite{JPW68} were
considered. It should be recognized a posteriori that those very first
computer studies attacked the problem from the most difficult side. In fact,
even before the effect of disorder was fully understood in harmonic chains,
they studied systems where anharmonicity and disorder are simultaneously 
present. Nevertheless, those early works have the merit to have
showed how the interplay of the two ingredients can lead to unexpected results
that, in our opinion, are still far from being fully understood. Indeed,
Ref.\cite{PRV67} revealed that the simple perturbative picture in which
anharmonicity and impurities provide two independent (and thus additive)
scattering mechanisms does not hold. More precisely, the Authors found even 
cases in which anharmonicity {\it enhances} thermal conductivity. A qualitative
explanation was put forward by claiming that anharmonic coupling induces an
energy exchange between the localized modes, thus leading to an increase of the
heat flux. Additional questions that have been investigated were the effect of
disorder on the temperature field and the concentration of impurities. Besides
the obvious finding that disorder reduces the value of heat conductivity (for
fixed finite-chain length), it was noticed an asymmetric behavior between the
case of a few heavy atoms randomly added to an otherwise homogeneous
light-atom chain and its converse. The smaller values of the conductivity
observed in the former cases were traced back to the larger number of localized
modes \cite{PRV67}.

Several studies of lattices under energy fluxes as well as attempts of
designing easy-to-simulate models followed these first studies. The reader is
referred to Ref.~\cite{LLP02} for a more detailed account. Among others it is
worth mentioning the example of the so-called ding-a-ling model~\cite{CFVV84}
that was the first example where the validity of the Fourier's law was
convincingly shown. Besides this neat instance (that actually belongs to a
different class of 1d chains which include the interactions with an external
substrate~\cite{Hu}), many (sometimes contradicting) results and interpretations
appeared in the literature. In retrospective, these difficulties can be traced
back to the presence of very strong correlations and slow dynamics that was
unexpected for such simple models and that could not be tackled due to
computational limits. In the next sections, we illustrate  how and why the
Fourier's law breaks down in the 1d FPU model.

\section{The quasi-integrable limit}

A remarkable property of models like~(\ref{FPU}) is the existence of two
distinct dynamical regimes for the relaxation dynamics. For the FPU--model, the
existence of an {\it energy threshold} was identified numerically by Bocchieri
et al.~\cite{Bocchieri} and confirmed by the resonance--overlap criterion
proposed by Chirikov and coworkers~\cite{ChIzTa}. Further numerical experiments
(see Ref.~\cite{Casetti} and reference therein) showed that
there exists a value of the energy density $e_c$ below which almost--regular
behavior significantly slows down the relaxation. This originates from the fact
that, although primary resonances do not overlap, higher order resonances can
do, yielding a slower evolution in phase space on a time scale that is
inversely proportional to a power of the energy density~\cite{DeLiRu}.
Conversely, above $e_c$, equipartition is rapidly reached during a typical
simulation run. For this reasons, $e_c$ has been termed {\it strong
stochasticity threshold}~\cite{Pettini}. Although the above analysis was mainly
focused on the FPU--$\beta$ model, it has been shown that such a threshold
generically exists for several lattice models in 1 and 2d~\cite{Casetti}. 

A related finding is the dependence of the maximal Lyapunov exponent $\lambda$
on the energy density. An analytic estimate of $\lambda$ for the FPU--$\beta$
problem has been also provided~\cite{CaLiPe}
\begin{equation}
\lambda(e) \sim
\left\{\begin{array}{ll}
e^2 &\quad {\rm for} \enskip e < e_c.\\
e^{1/4} &\quad {\rm for} \enskip e > e_c;\\
\end{array}\right.
\label{lyeps}  
\end{equation}
This implies that the ``correlation'' time due to the presence of chaotic
instabilities (i.e. the inverse of $\lambda$) may become exceedingly large for
small enough values of $e$. 

\begin{figure}[ht!]
\includegraphics[clip,width=6.5cm]{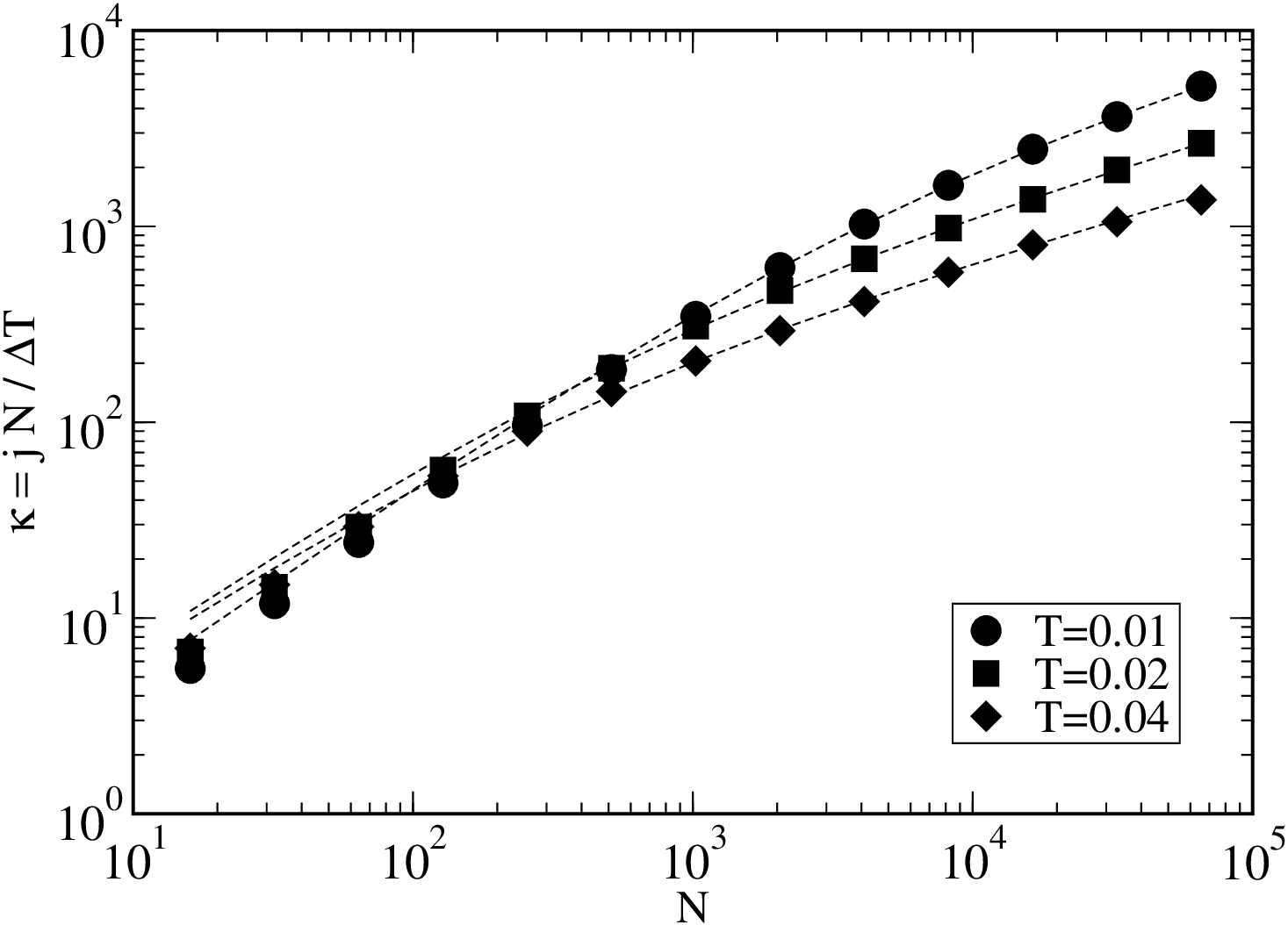}
\caption{
Finite-size conductivity for the FPU-$\beta$ model below the strong 
stochasticity threshold for three different choices of the
boundary temperatures $T_\pm=T\pm\Delta T/2$; $\Delta T=0.01$.
Nos\`e--Hoover thermostat with response times fixed to $0.5$;
each point results from an average over a single trajectory of
about $10^6$ time units. Dashed lines are the best fits 
according to Eq.~(\ref{fitte}).
} 
\label{fig:cond}
\end{figure}

\begin{figure}[ht!]
\includegraphics[clip,width=6.5cm]{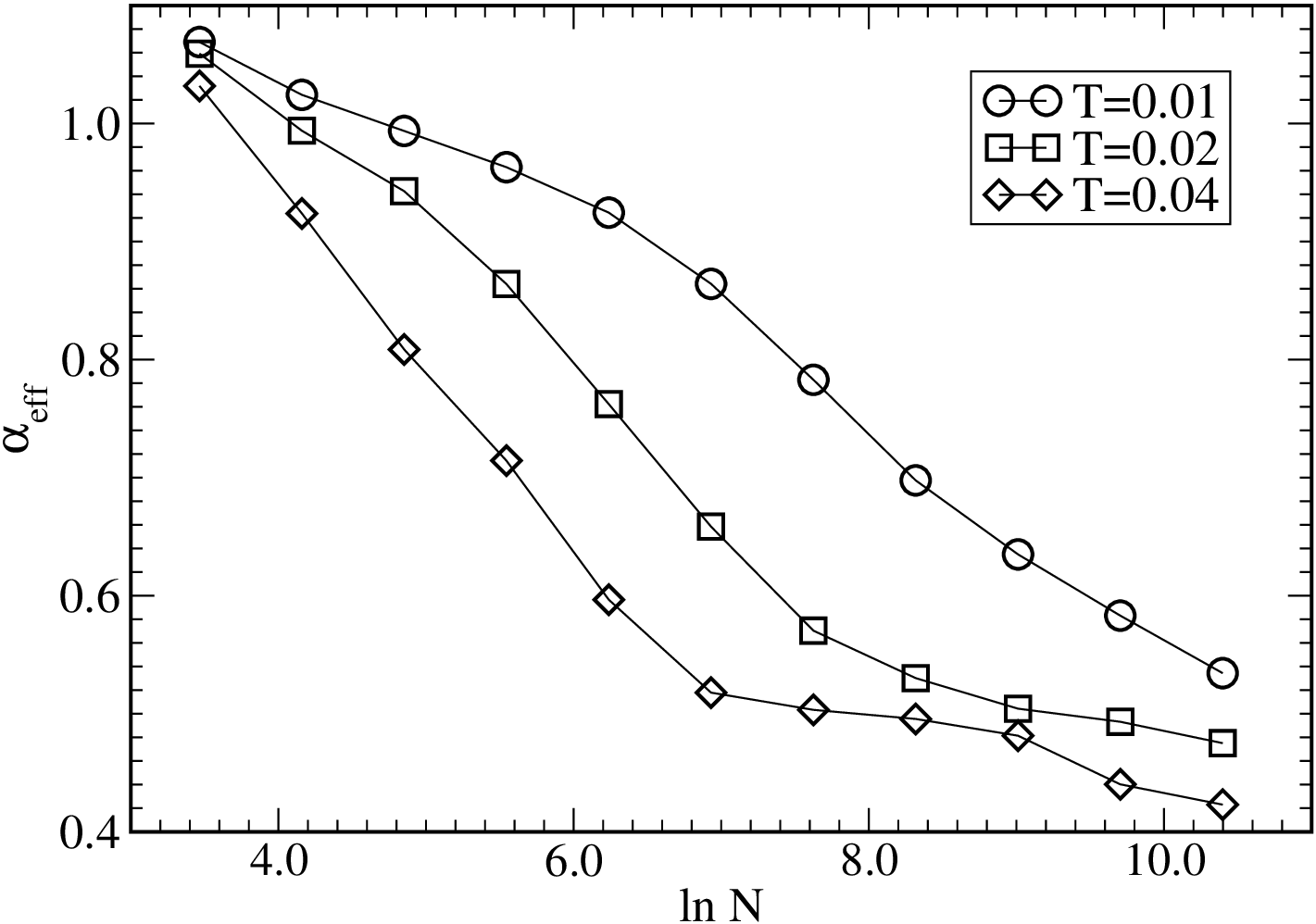}
\caption{
The effective exponent $\alpha_{\rm eff}$ of the finite-size
conductivity for the FPU--$\beta$ model below the strong stochasticity threshold.
Points are obtained by evaluating the centered differences from 
data in the previous figure.
} 
\label{fig:dlog}
\end{figure}

The existence of long--lasting transients and the fact that the
Lyapunov time--scale becomes very large both suggest that also stationary
transport properties below $e_c$ may be strongly affected by the 
quasi--integrability of the dynamics. To investigate this features, we
simulated the FPU-$\beta$ chain in contact with reservoirs at different
temperatures. For computational convenience, the parameters $m$, $\omega$
and $\beta$ have been fixed to unity so that the only relevant physical 
parameter is the energy per particle $e$. With this choice the threshold energy
is $e_c\simeq 0.1$~\cite{Pettini}. The average temperature $T=(T_++T_-)/2$ has
been chosen to yield an average internal energy of the chain below $e_c$. We
have used the Nos\'e--Hoover thermostats described in detail in
Ref.~\cite{LLP02}.
In order to fasten the convergence towards the stationary state, the initial
conditions have been generated by thermostatting each particle to yield a
linear temperature profile. This method is very efficient,
especially for long chains, when bulk thermalization may be significantly slow.
Simulations of the FPU--$\beta$ model with chains of length
up to $65536$ sites and free boundary conditions exhibit a monotonous increase
of the finite--size conductivity (see Fig.~\ref{fig:cond}). The growth is 
linear at small lengths and crosses over to a slower increase.
This is best seen as a systematic decrease of the effective exponent 
\be  
\alpha_{\rm eff}(N) \;=\;
\frac{d \ln \kappa}{d \ln N} ,  
\label{aeff} 
\ee 
from $\alpha_{\rm eff}\simeq 1$ to a $\alpha_{\rm eff}\simeq 0.5$ 
(see Fig.~\ref{fig:dlog}). 

The data reported in Fig.~\ref{fig:cond} does not a priori exclude the 
possibility of a slow convergence to a constant value. Since we rather expect
a power law divergence (see next section) we tentatively fitted the 
conductivity data with a form (see also~\cite{AK01})
\begin{equation}
{1\over \kappa(N)} \;=\; {a\over N} \;+\;{b\over N^\alpha}
\label{fitte}
\end{equation}
The results of the nonlinear fit are shown as solid lines in 
Fig.~\ref{fig:cond} and the fitted parameter are reported in Table~\ref{tab1}. 
Notice that the exponent $\alpha$ is almost independent of the temperature.

\begin{table}
\caption{Fitting parameteters,  (Eq.~\ref{fitte}) 
\label{tab1}}
\begin{ruledtabular}
\begin{tabular}{cccc}
$T$             &  $a$      & $b$  & $\alpha$   \\
\hline
0.01            &   2.00   &   0.013 &  0.400  \\
0.02            &   1.28   &   0.039 &  0.423  \\
0.04            &   1.31   &   0.059 &  0.404 \\
\end{tabular}
\end{ruledtabular}
\end{table}

\begin{figure}[ht!]
\includegraphics[clip,width=6.5cm]{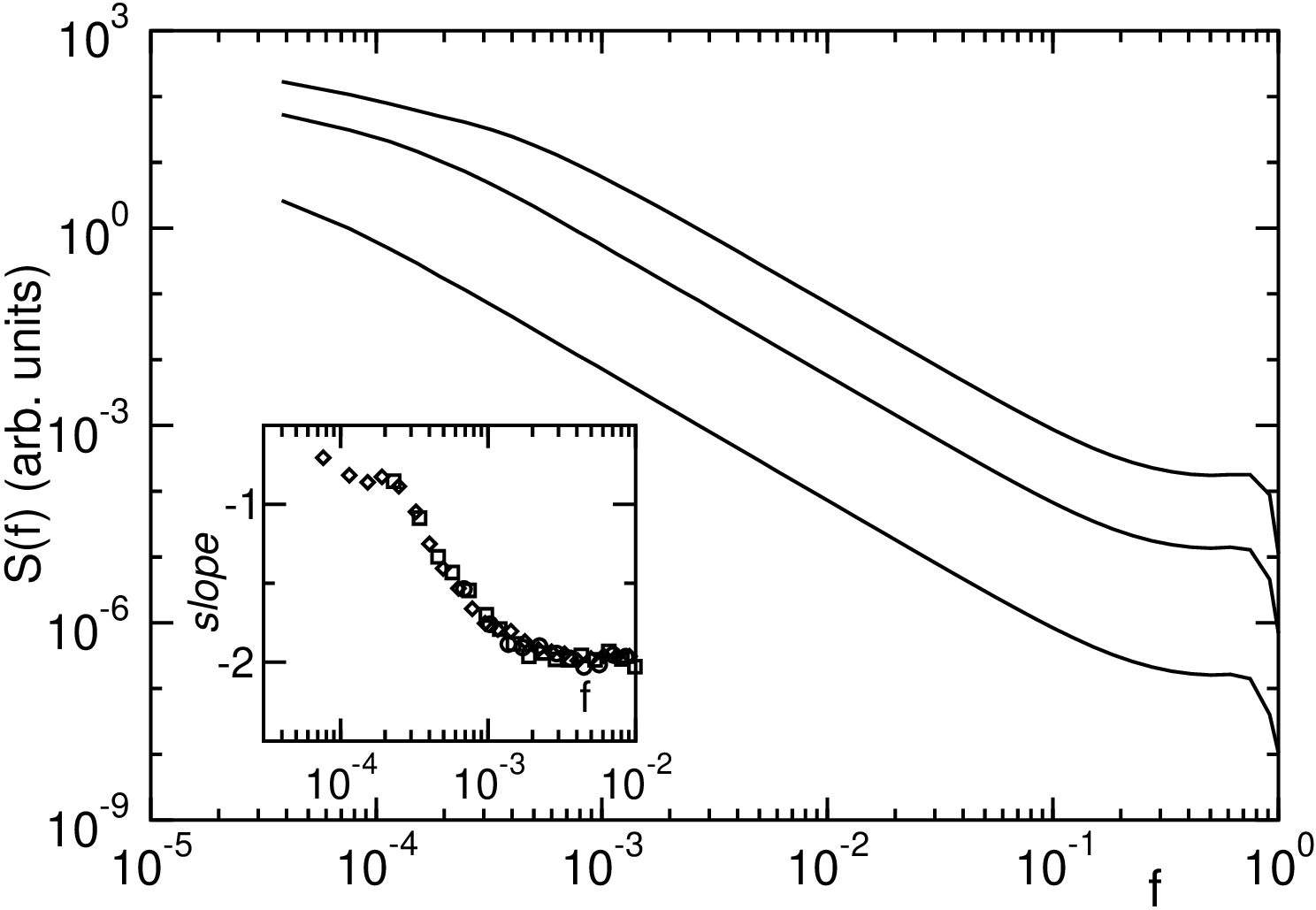}
\caption{
Power spectra of the flux $J$ as defined in Eq.~(\ref{jsolid}) for 
the quartic FPU--$\beta$ model with $N=2048$ (solid).
Data are averaged over about 2500 random initial conditions. To
minimize statistical fluctuations, a binning of the data over
contiguous frequency intervals has been performed. The curves 
are for energies $e=0.01$, $e=0.02$, $e=0.04$ (bottom to top) and
have been vertically shifted for clarity. In the inset, the logarithmic
derivative is reported versus the frequency, after a suitable rescaling of the
latter quantity. Circles, squares and diamonds refer to $e=0.01$, 0.02, and
0.04, respectively.} 
\label{fig:sfpu}
\end{figure}

Transport properties can be analyzed by computing the power spectrum $S(f)$ of
the heat current $J$ at equilibrium.  We choose to work in the microcanonical
ensemble and integrated the equations of motion with periodic boundary
conditions and zero total momentum  with a fourth--order symplectic algorithm
\cite{MA92}. The spectra obtained for three different energies roughly
corresponding to the average temperatures in Fig.~\ref{fig:cond} are reported
in Fig.~\ref{fig:sfpu}. By comparing the results obtained for different chain
lengths, it is possible to conclude that finite-size corrections are negligible in the
considered spectral range. From the inset, where the logarithmic derivative 
\begin{equation}
\delta_{\rm eff}(f) \;=\; \frac{d \ln S}{d \ln f} \quad.
\label{deff}
\end{equation}
is reported versus the frequency $f$, one can notice that the Lorenzian tail
observable at high frequencies starts to cross over towards a regime
characterized by a weaker divergence and presumably controlled by the same
mechanisms operating in the strong chaotic regime (see next section). Although
one can see some evidence of a second plateau only for the largest temperature
($T=0.04$ -- see diamonds in the inset of Fig.~\ref{fig:sfpu}), it is possible
to investigate how the crossover time $t_c$ increases upon decreasing the
energy density by suitably rescaling the frequency axis. As a result, it turns
out that $t_c$ is roughly proportional to $e^{-1.6}$, 
which is not too far from the divergence $e^{-2}$ of the Lyapunov time--scale 
in the weakly chaotic regime (see Eq.~(\ref{lyeps})). It is thus
reasonable to infer that the dynamical mechanisms underlying the slow 
relaxation and the anomalous conduction are basically the same 
at such low temperatures.

\section{The strong--chaos regime}

In the previous section we saw how the almost--regular features of the 
dynamics may give rise to a breakdown of a macroscopic transport law (the
Fourier's law in this case). On the basis of the previous discussion, it might
be surmised that above the strong stochasticity threshold such anomalies should
somehow disappear as the dynamics becomes more ``mixing" and the relaxation
times shorter. It thus came as a surprise~\cite{KM93} when the anomalous
transport features  of the FPU chain were actually discovered to persist also
in this  regime~\cite{LLP97}. The origin of this anomalous behavior is
twofold.  The first cause is the {\it reduced dimensionality}. Indeed, strong
spatial constraints can significantly alter transport  properties: the response
to external forces depends on statistical fluctuations which, in turn,
crucially depend on the system dimensionality $D$. This is very much
reminiscent of the problem of long--time tails in fluids \cite{PR75} where, for
$D\le2$, transport coefficients may {\it not exist at all}. The second
necessary condition is {\it momentum conservation} i.e. the existence of
long-wavelength (Goldstone) modes that are propagating and very weakly damped
($\tau_k$ diverges for  small $k$). This latter condition is necessary for
these anomalies to occur and means that no external (i.e. substrate) forces
must be present. This is precisely the case of the cited ding-a-ling
model~\cite{CFVV84} and of other models in the same class (the nonlinear
Klein-Gordon chain for example~\cite{Hu}). The only remarkable exception to
this is the coupled rotor chain where, however, different mechanisms are at
work~\cite{GLPV00}.

Anomalous behaviour means both a nonintegrable algebraic decay of equilibrium
correlations of the heat current $J(t)$ (the Green-Kubo integrand) at large 
times $t\to \infty$ and a divergence of the finite-size conductivity 
$\kappa(L)$ in the $L\to \infty$ limit.  
As a results of a series of simulation studies \cite{LLP02}, it can
be stated that for 1d lattice models one finds
\be 
\kappa(L) \propto L^\alpha   \quad, \qquad
\langle J(t)J(0)\rangle \propto t^{-(1+\delta)}  \quad ,  
\label{anomal}
\ee
where $\alpha >0$, $-1<\delta < 0$, and  $\langle \,\, \rangle$ is the
equilibrium average. For small applied gradients, linear-response theory allows
establishing a connection between the two exponents. By assuming that
$\kappa(L)$ can be estimated by cutting-off the integral in the Green-Kubo
formula at the ``transit time'' $L/v$ ($v$ being some propagation velocity of
excitations), one obtains $\kappa \propto L^{-\delta}$ i.e. $\alpha=-\delta$. 

It is thus natural to argue about the universality of the exponent $\alpha$. 
On the one hand, two independent theoretical approaches, self-consistent
mode-coupling approximation~\cite{E91,LLP98,WL04} and kinetic
theory~\cite{P03}, yield  $\alpha=2/5$. On the other hand, a renormalization
group calculation on the stochastic hydrodynamic equations for a 1d
fluid~\cite{NR02} gives $\alpha=1/3$. Validation of one of the theories is
still under debate, and the existence of crossovers among different scaling
regimes has been observed~\cite{WL04}. The available numerical data for
$\alpha$ range from 0.25 to 0.44~\cite{LLP02,LLP03}. As a word of caution, it
must be stressed that a numerical estimates are indeed challenging. Even in the
most favorable case of  computationally efficient models, as the $1D$ gas of
hard-point particles with alternating masses~\cite{H99}, finite--size
corrections to scaling are sizeable. As a matter of fact, estimates of 
$\alpha$ as diverse as 0.33 ~\cite{GNY02,LWZ02} and 0.25~\cite{CP03} for
comparable parameter choices have been reported.  Although in this latter
instance the anomaly is possibly due to the lack of microscopic
chaos~\cite{GNY02}, it is a generic fact that sensible results require reaching
the limits of present computing resources. For instance, in the case of the FPU
chain, the best estimate sofar (see below) required simulations of up ${\cal
O}(10^4)$ particles and ${\cal O}(10^8)$ integration steps (plus ensemble
averaging)~\cite{LLP03}.

Since we are interested in the long--wavelength and small--frequency behavior,
it is convenient to consider a highly nonlinear model in the hope that the
asymptotic regime sets in over shorter time and space scales. Moreover, it is
advisable to work with a computationally simple expression of the force.
A reasonable compromise is the ``infinite temperature" FPU
model~\cite{LLP03}
\begin{equation}
V(z) \;=\;  {1\over 4}\, z^4 \quad .
\label{fpu4}
\end{equation}
This model has no free parameters: since the potential expression is homogeneous,
the dynamics is invariant under coordinate rescaling, so that the energy per
particle $e$ can be set, without loss of generality, equal to 1.

\begin{figure}[ht!]
\includegraphics[clip,width=6.5cm]{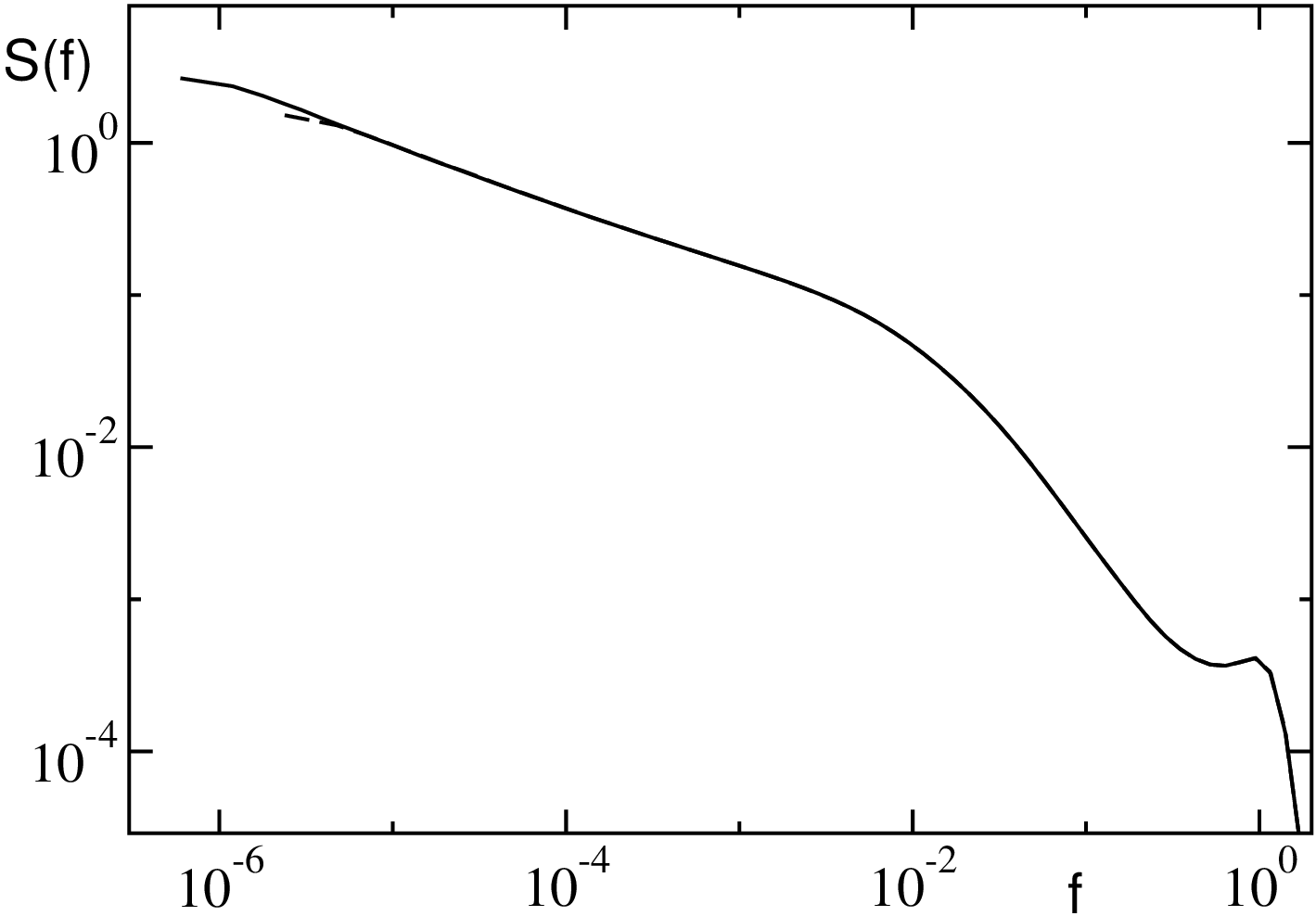}
\caption{
Power spectra of the flux $J$ as defined in Eq.~(\ref{jsolid}) for 
the quartic FPU model (\ref{fpu4}) with $N=2048$ (solid)
and 1024 (dashed). 
Data are averaged over 30,000 random initial conditions. To
minimize statistical fluctuations, a binning of the data over
contiguous frequency intervals has been performed.  } 
\label{fig1}
\end{figure}

The power spectra $S(f)$ of $J$ are reported in Fig.~\ref{fig1}. The long--time
tail (\ref{anomal}) corresponds to a power--law divergence $f^\delta$ in the
low-$f$ region. By comparing the results obtained for different numbers of
particles, one can clearly see that finite-size corrections are negligible
above a size-dependent frequency $f_c(N)$.  By fitting the data in the scaling
range $[f_c(N),f_s]$, where $f_s \simeq 10^{-3}$ , we find $\delta=-0.39(6)$.
These values are consistent with previous, less-accurate, findings for similar
models, such as the standard FPU \cite{LLP98,GY02} and the diatomic Toda
\cite{H99} chains, thus confirming the expectation that they all belong to the
same universality class.

In order to perform a more stringent test of the scaling behavior, we again 
evaluateid the logarithmic derivative (\ref{deff})
for different frequencies. Since finite-size effects are responsible for the
saturation of $S(f)$ when $f \to 0$, $f_c(N)$ can be identified (see
Fig.~\ref{fig1}) as the frequency below which $\delta_{\rm eff}$ starts growing
towards zero. Above $f_c$, the quality of our numerical data allows revealing a
slow but systematic decrease of $\delta_{\rm eff}$ upon decreasing $f$, which
approaches $-0.44$, a value that is incompatible not only with the
renormalization-group prediction of Ref.~\cite{NR02}, but also with the result
of mode--coupling ~\cite{E91,LLP98} and kinetic \cite{P03} theories.
Furthermore, convergence seems not fully achieved in the accessible frequency
range.

\begin{figure}[ht!]
\includegraphics[clip,width=6.5cm]{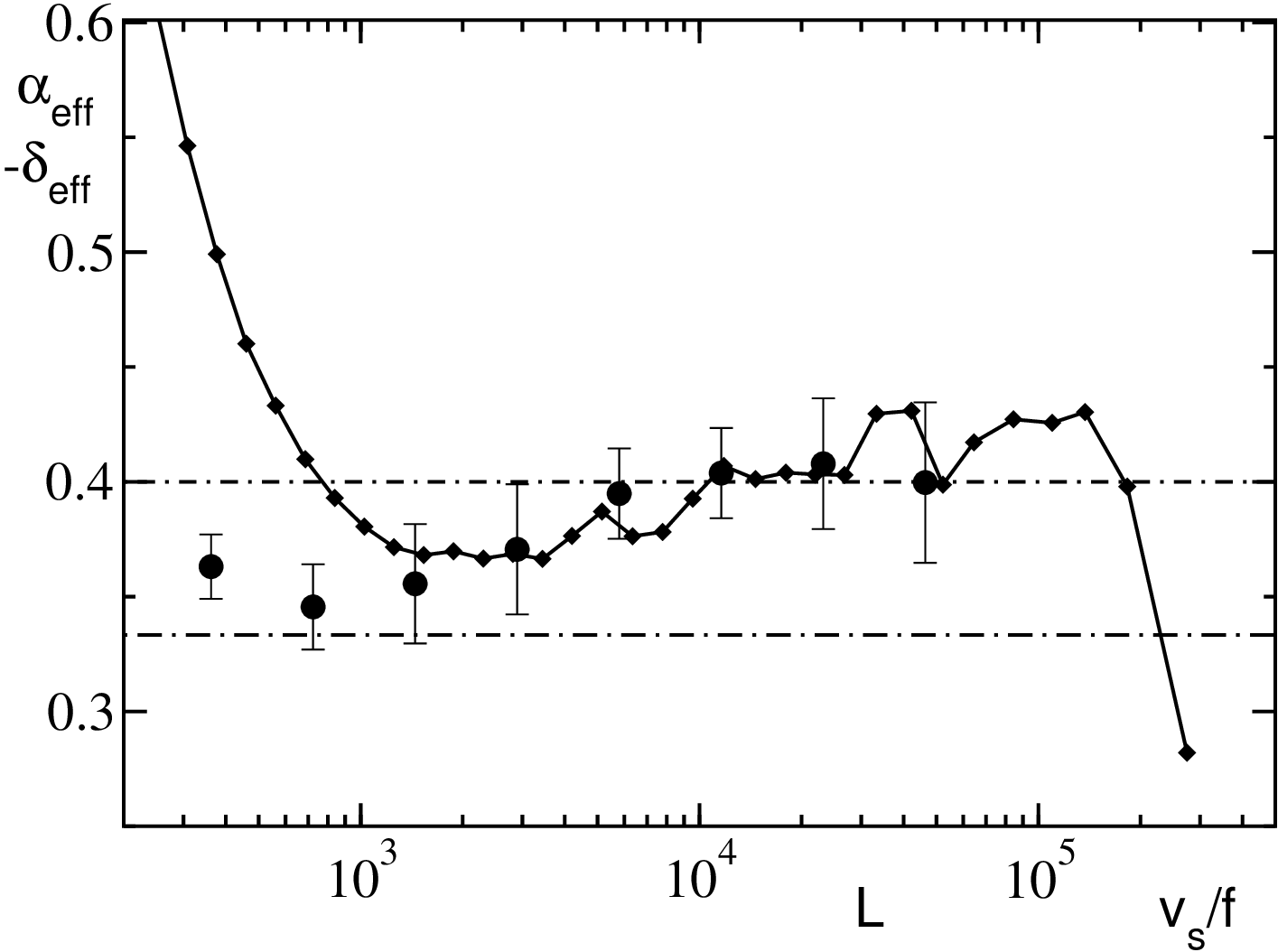}
\caption{
Quartic FPU model: the effective exponent $\alpha_{\rm eff}$ of the finite-size
conductivity for $T_+=1.2, T_-=0.8$ (full dots), compared with the results
($-\delta_{\rm eff}$) of equilibrium simulations. The two horizontal lines
correspond to the theoretical predictions, 1/3 and 2/5.} 
\label{fig:scaling}
\end{figure}

To check the consistency of equilibrium and nonequilibrium simulations,  one
can assume, following the argument exposed below Eq.~(\ref{anomal}),  that the
finite--size conductivity $\kappa(L)$ (with $L=Na$) is determined by
correlations up to time $\tau = L/v_s$, where $v_s$ is the sound velocity. This
means that the frequency $f$ can be turned into a length $L=v_s/f$. It might be
argued that  the absence of a quadratic term in (\ref{fpu4}) prevents a
straightforward  definition of such a velocity in the $T=0$ limit;
nevertheless, it has been  shown~\cite{AC03} that an effective phonon
dispersion relation at finite energy density can be evaluated for model
(\ref{fpu4}), yielding $v_s = 1.308$ at  $e=1$. Using this value, we can
ascertain that, at least for $N>1000$, there is an excellent agreement between
the two approaches (see again Fig.~\ref{fig:scaling}).

\begin{figure}[ht!]
\includegraphics[clip,width=6.5cm]{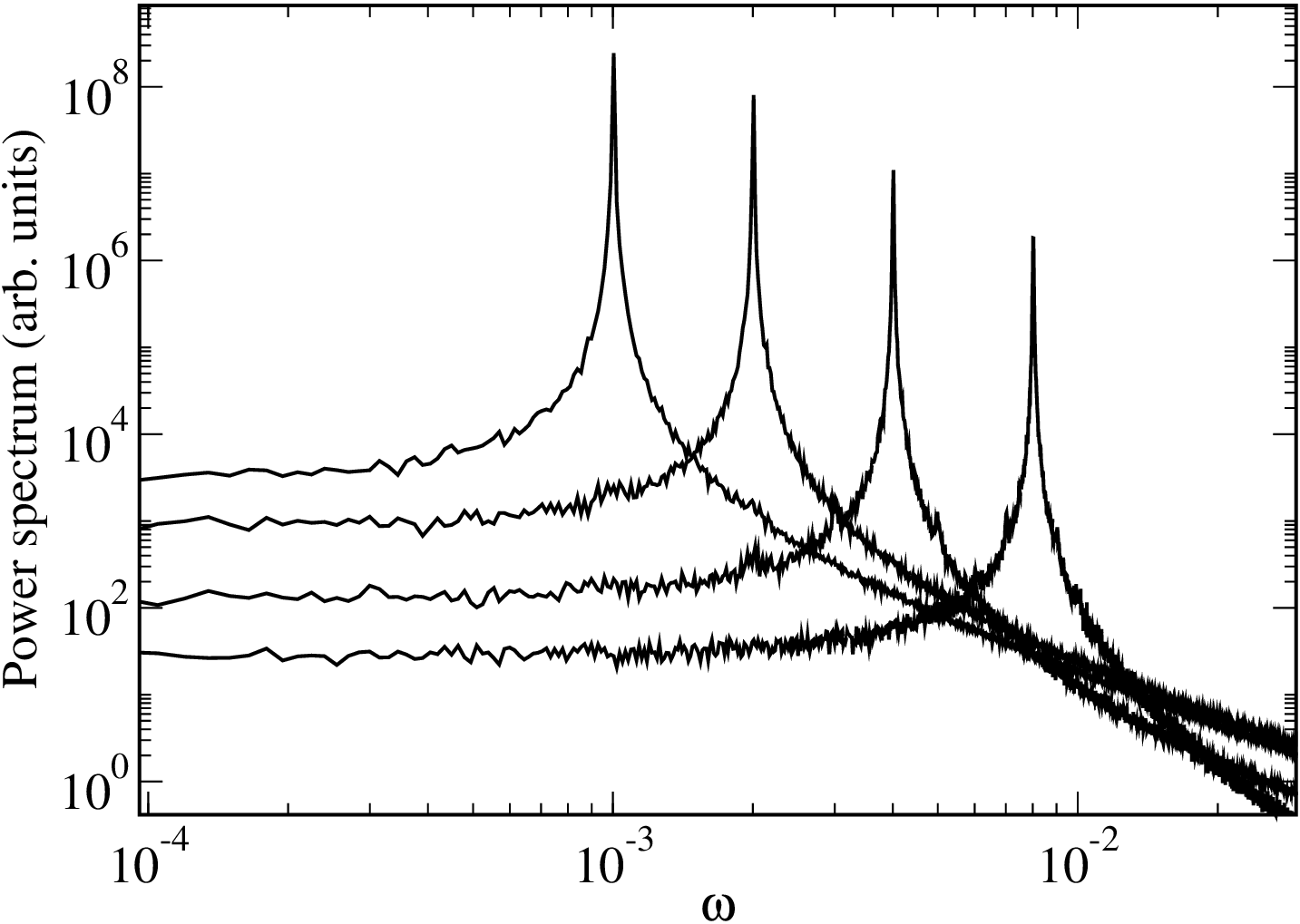}
\caption{
Structure factors for the quartic FPU model (\ref{fpu4}) with $N=8192$
(solid) for the modes of indexes $k=1, 2, 4, 8$ (left to right). 
Microcanonical simulations are performed for the energy density $e=1$ and 
averaged over an ensemble of about 200 initial conditions. 
} 
\label{fig:spectra}
\end{figure}

\begin{figure}[h]
\begin{center}
\includegraphics[clip,width=6.5cm]{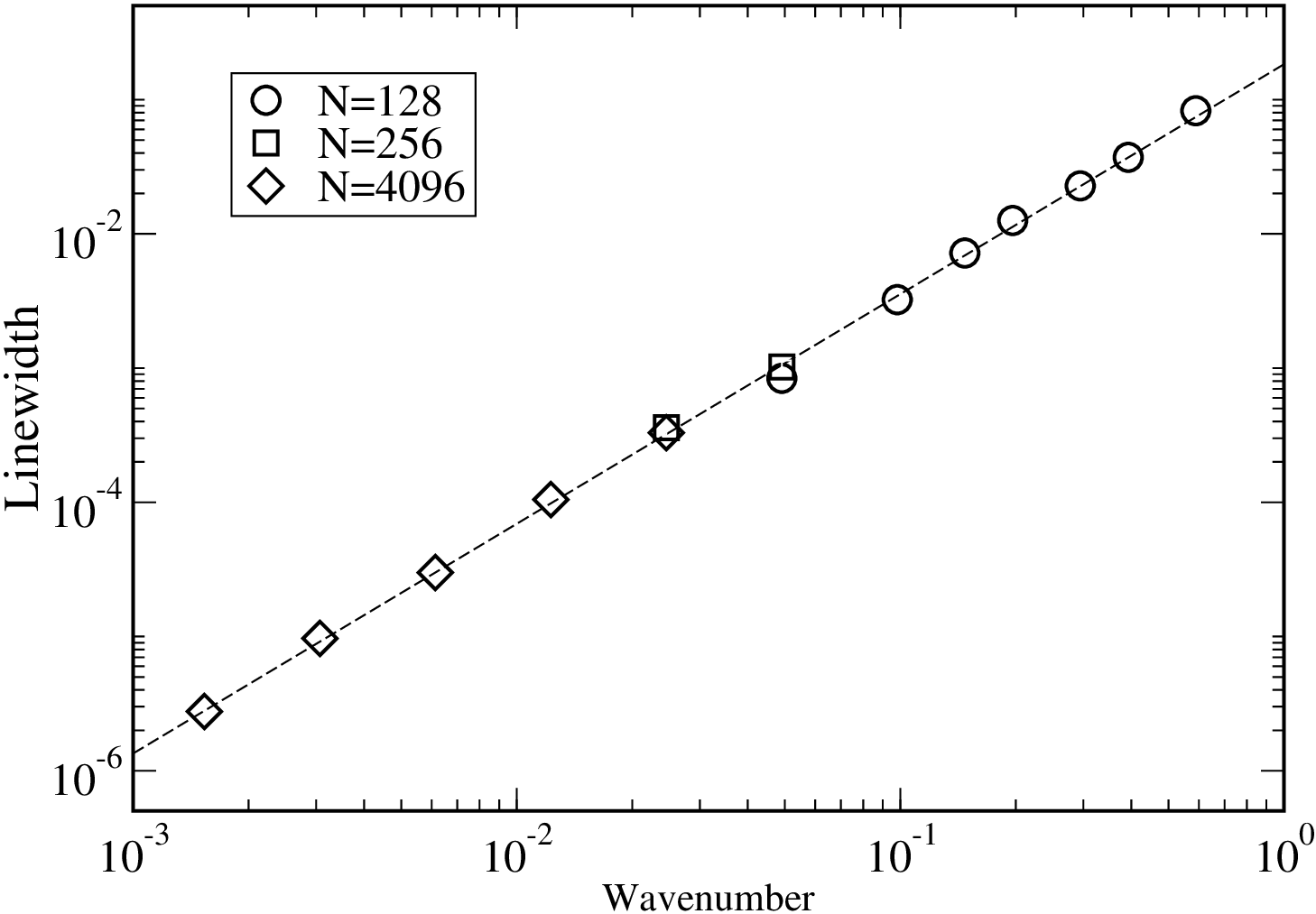}
\caption{The wavenumber dependence of the linewidths 
for the quartic FPU model (\ref{fpu4}). The linewidths are obtained 
by fitting the peaks of the spectra for different chain lengths 
with a lorenzian lineshape. The dashed line is a power--law fit
yielding an exponent $1.677\pm 0.025$.}
\label{fig:gamma4}
\end{center}
\end{figure}

Another way to detect hydrodynamic anomalies is to measure the equilibrium 
structure factors, namely the averaged power spectra of the mode amplitudes
\begin{equation}
Q_k \;=\;{1\over \sqrt{N}} \sum_{n=1}^N \, q_n \, e^{-i{2\pi k \over N}n} 
\label{Qk}
\end{equation} 
The spectra display a phonon--like peak at a frequency which is
in excellent agreement with the one computed in Ref.~\cite{AC03}
(see Fig.~\ref{fig:spectra}): the 
estimated sound speed is $v_s=1.34\pm 0.04$, see above. The peaks' linewidths
provide a measure of the inverse of the relaxation times; they are 
found to scale as a power of the wavenumber (see Fig.~\ref{fig:gamma4})
with an exponenent which is very close to the estimate $5/3$ based 
on arguments of mode-coupling theory~\cite{E91}. Notice that in the 
standard case one would rather expect a quadratic law, corresponding 
to the usual friction term of elasticity theory. 

\section{Modal analysis}

Conductivity properties are mostly investigated in real space because one is
interested in the dependence of the heat flux on the system size. Nevertheless,
the renormalization group analysis has teached us that a (spatial) Fourier
analysis can be very useful especially to understand anomalous scaling
behaviours. Additionally, in the case of integrable systems, a modal analysis
allows solving the problem by decomposing it into many independent parts; last
but not least the effect of boundary conditions can be better appreciated. In
spite of such advantages, very few numerical studies have been devoted to
investigating heat conductivity in Fourier space for nonlinear 
chains~\cite{FVMMC97,LLP02}.
Here in the following, we summarize the current understanding and discuss the
open problems with the help of numerical simulations to have some hints
about the expected scenario.

As proposed in Ref.~\cite{FVMMC97}, it is convenient to start from the formal
energy balance
\begin{equation}
  J_k^+ + J_k^- + J_k^{nl} = 0  ,
\label{enebalance}
\end{equation}
where $J_k^+$, $J_k^-$, and $J_k^{nl}$ denote the energy exchanged (per unit
time) by the $k$th mode with the hot, cold heat bath, and with the other modes,
respectively. By summing over the index $k$, one finds that
\begin{equation}
  J^+ + J^- = 0  ,
\label{balantot}
\end{equation}
where $J^{\pm}$ is the total energy flowing from (to) the hot (cold) bath;
in fact $\sum_k J_k^{nl}=0$, since the energy of the system is, on the average,
constant. As a result, the total heat flux $J$ can be, e.g., decomposed in
modal contributions,
\begin{equation}
  J =  \sum_k J_k^+
\end{equation}
So far, this is just a formal statement. In order to make it operative, it is
necessary to give an explicit definition of the modal fluxes. Using the 
definition (\ref{Qk}), one can show 
that the dynamics of the $k$th mode is described by the following equation
\begin{equation}
   \ddot Q_k + \omega_k^2 Q_k + \phi_k^{nl} + \phi_k^+ + \phi_k^- = 0,
\label{modeq}
\end{equation}
where $\omega_k$ is the frequency of the $k$ mode (which follows from an energy
dependent renormalization of the harmonic frequency), while $\phi_k^{nl}$ is
the effective force due to the interactions with all other modes, while
$\phi_k^\pm$ account for the interaction with the thermal baths.

Upon multiplying Eq.~(\ref{modeq}) by $\dot Q_k$, and introducing the modal energy
$E_k = (\dot Q_k^2 + \omega_k^2 Q_k^2)/2$, it follows that
\begin{equation}
  0 = \langle \dot E_k \rangle = \langle \phi_k^{nl} \dot Q_k \rangle
    + \langle \phi_k^+ \dot Q_k \rangle + \langle \phi_k^- \dot Q_k \rangle 
\end{equation} 
where $\langle \cdot \rangle$ denotes a time average and the first equality
is a trivial consequence of stationarity. Since we are assuming that the lattice
spacing is $a=1$, it is now clear that the three terms in the r.h.s. of the
above equations coincide with the modal fluxes $J_k^{nl}$, $J_k^+$, and $J_k^-$,
respectively.

All such contributions can be easily computed when the interaction with the
thermal bath amounts to instantaneous stochastic collisions occurring to
either the first or last particle. In such cases, the fluxes $J_k^\pm$
can be easily computed by a mode expansion of the energy locally
exchanged in each collision, while $J_k^{nl}$ can be obtained from the
energy variation in between collisions.

In a linear chain, the eigenmodes are independent of each other, so that
the term $J_k^{nl}$ vanishes. In such conditions, the effective equation
of the $k$th wavevector reduces to
\begin{equation}
 \ddot Q_k + \omega_k^2 Q_k + \gamma_k^+\dot Q_k  + \gamma_k^-\dot Q_k +
  \xi^+ + \xi^- = 0,
\end{equation}
where $\gamma_k^\pm$ gauges the interaction with the thermal bath while 
$\xi_k^\pm$ is a Gaussian white noise, whose diffusion constant is,
according to the fluctuation-dissipation theorem,
$D_k^\pm = k_B T^\pm \gamma_k$.
The above equation is nothing but a Langevin
equation describing an ``oscillator" with a dissipation
$\gamma_k^+ + \gamma_k^-$ and an effective noise corresponding to
the intermediate temperature 
$T =(\gamma_k^+ T^+ + \gamma_k^-T^-)/(\gamma_k^+ + \gamma_k^-)$.
Although the average
energy of the oscillator neither increases nor decreases, the energy exchanged
per unit time with each bath 
\begin{equation}
  J_k^\pm = -\gamma_k^k \langle \dot q_k^2 \rangle -
             \langle \xi^+ \dot q_k^2 \rangle
\end{equation}
is different from zero. 

Let us illustrate this under the simplifying but
otherwise general assumption of equal coupling strength with the heat baths
($\gamma_k^+ = \gamma_k^- = \gamma_k$). In fact, in this case,
$T=T_a=(T^+ + T^-)/2$ so that the energy exchanged with the hot heat bath is
$J_k^+ =-2\gamma_k T + 2\gamma_k T^+ = \gamma_k(T^+-T^-)$, i.e.
$J_k \approx \gamma_k$. Conductivity properties can thus be understood from
the coupling strength of the eigennodes. In harmonic systems, it is well
known that $\gamma_k$ is proportional to the square amplitude of the $k$th
eigenmode at the chain end~\cite{LLP02}. For fixed and free b.c.,
$\gamma_k \approx k^2/N^3$, $\gamma_k \approx 1/N$, respectively (for the sake
of simplicity, we neglect the nonlinear dependence on $k$, that is important
only at large wavenumbers). Since the total heat flux is obtained by summing all
contributions up to $k=N$, it follows that $J$ is a quantity of order one
independently of the b.c..

The scenario becomes very different if some disorder is added, because the
exponential localization of the eigenmodes makes their coupling with the heat
baths basically negligible. In such conditions, only the first $\sqrt{N}$ modes
are sufficiently extended to appreciably contribute to the heat flux.
For free b.c., the total flux being the sum of the contribution is on the order
of $1/\sqrt{N}$ which corresponds to a square root divergence of the
condcutivity $\kappa$. On the contrary,
for fixed b.c., the sum of $\gamma_k = k^2/N^3$ yields $J \approx 1/N^{3/2}$,
i.e. a vanishing conductivity. In other words, in the presence of disorder,
a change of b.c. may turn a ``heat superconductor" into a very good insulator!.

The question then arises of how the scenario modifies in the presence of
nonlinearities. Past investigations of the FPU models clearly indicate
that in the absence of interactions with thermal baths, the dynamics of a
Fourier mode with small $k$ is well described by the Langevin equation
\cite{L98}
\begin{equation}
\ddot Q_k + \omega_k^2 Q_k + \gamma_k^{nl} \dot Q_k + \xi^{nl} = 0 \quad. 
\end{equation}
According to mode-coupling theory, $\gamma_k \approx (k/N)^{5/3}$
an estimate that is very close to the numerical values 
(see also Fig.~\ref{fig:gamma4}).
It thus appear natural to assume that in the presence of heat baths, the
above equation modifies to
\begin{equation}
\ddot Q_k + \omega_k^2 Q_k + (\gamma_k^{nl} + 2\gamma_k)\dot Q_k + 
   \xi^{nl}  + \xi^+ + \xi^-
\label{modgeneq}
\end{equation}
although we anticipate that the effective value of the coupling $\gamma_k$ with
the heat bath cannot be the same as in the harmonic case. We start addressing
the problem of the effective temperature of the $k$th mode. By repeating the
same arguments that lead to predict a temperature $T_a$ in the harmonic case,
we obtain now 
$T_k= [2\gamma_k T_a + \gamma_k^{nl}T_k^{nl}]/(2\gamma_k + \gamma_k^{nl})$
where we write $T_k$ to emphasize a possible dependence of the temperature on
the wavenumber and where $T_k^{nl}$ is basically unknown, and it is
expected to range between $T^-$ and $T^+$. For free b.c.,
$\gamma_k \approx 1/N$, thus implying that there exists a critical value
$k_c \approx N^{2/5}$ below which $\gamma_k^{nl}$ is negligible with respect to
$\gamma_k$ and viceversa above it. As a result, below $k_c$, we expect
$T_k = T_a$, while above, there should be a crossover towards the unknown value
$T_k^{nl}$. Simulations performed in chains of different lengths with
$T^+ = 10$ and $T^- = 8$ are trivially in agreement with this scenario.
Indeed, $T_k$ determined by evaluating the kinetic temperature of the $k$th 
mode is constantly equal to $T_a=9$. This implies that in spite of the
nonlinearity in the profile, modal temperatures behave exactly as in the
harmonic case (and also as when the Fourier law would apply). The same is true
also for free boundary conditions when, being $\gamma_k = k^2/N^3$, there does
not exist a value $k_c$, since external dissipations should be always
negligible.

\begin{figure}[tcb]
\centerline{\includegraphics[clip,width=7cm]{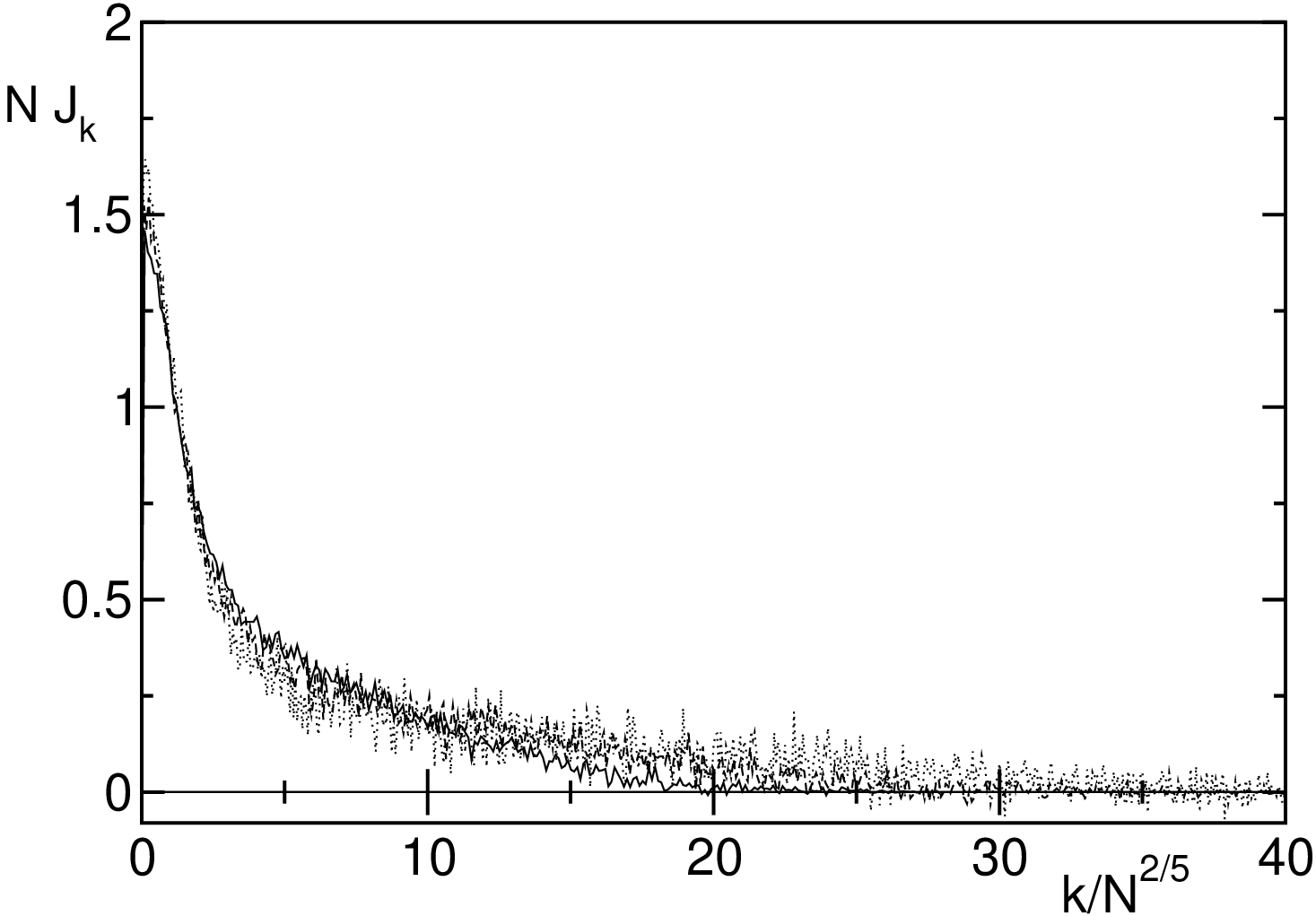}}
\caption{ \label{figfluxmod}
Modal flux $J_n$ (multiplied by the chain length $N$ versus the scaled
wavenumber $k/N^{2/5}$ for an FPU-$\beta$ lattice with free boundary
conditions. The solid, dashed and dotted lines refer to $N=256$, 512 and
1024, respectively. Averages are taken over a time span of $10^8$ units.}
\end{figure}

Let us now analyse the behaviour of modal fluxes. As the internal noise
temperature
adjusts itself to the arithmetic average of the heat-bath temperatures, the
flux $J_k^{nl}$ should vanish. This is confirmed by our simulations: the small
values of $J_k^{nl}$ that we have obtained can indeed be interpreted as
statistical fluctuations. Much more intriguing is the analysis of the true
energy fluxes $J_k^\pm$. With reference to free boundary conditions, the 
integral flux due to the modes with $k<k_c$ scales as $1/N^{3/5}$ and this
would be consistent with the direct computations of heat conductivity, were the
contribution of $J_k^\pm$ for $k>k_c$ negligible. Unfortunately, if we assume
that $\gamma_k$ keeps scaling as $1/N$ (like in the harmonic case) we are led to
conclude that all channels equally contribute to the heat flux with a consequent
linear divergence of the conductivity with the system size! We already know that
this is not the case, but the data reported in Fig.~\ref{figfluxmod} give us
some hints about possible explanations of this failure.
In the figure we report the modal flux multiplied by $N$ versus $k/N^{2/5}$
for three different chain lengths. The reasonably good overlap confirms the
hyphotesis that for $k<k_c \approx N^{2/5}$, the internal dissipation is
negligible and the system behaves like a harmonic chain. On the other hand, the
decrease of $J_k$ suggests that above $k_c$ the effective coupling with the
thermal baths is much smaller than $1/N$. Determining the dependence of
$\gamma_k$ on $k$ in this regime is, to our knowledge, an open problem.
Solving this problem is all the way more crucial in the context of fixed
boundary conditions where $\gamma_k$ must differ everywhere from the harmonic
chains. In fact, simulations publised elsewhere \cite{LLP02} indicate that, at
variance with the harmonic case, although $J_k$ vanishes for $k \to 0$, the
overall scaling behaviour remains the same as in the free b.c. case.

Finally, we wish to mention that the modal analysis can help to shed some light
on a futher open problem: the determination of the temperature profile. We have
seen that different models (such as harmonic and FPU chains) are characterized
by energy equipartion even out of equilibrium. This is true even though they
exhibit very different spatial profiles. It is quite natural to conjecture that
this is caused by different phase correlations among the various modes, but
working out a detailed explanation appears not to be a trivial task.

\section{The 2D FPU}

The presence of an energy threshold for the relaxation 
dynamics has been observed also in 2d lattice models of anharmonically
coupled oscillators. For instance, in \cite{BT} the authors 
investigated a square--lattice of oscillators coupled by a 
Lennard--Jones 6/12 potential
\be
V(r) = A\left[ (\frac{\sigma}{r})^{12} - 
(\frac{\sigma}{r})^6 \right]
\label{LJ}
\ee
where $r=|\vec{r}|$ is the modulus of the distance vector $\vec{r}$ 
between nearest-neighbor oscillators.
At energy density $e_c \approx 0.3$ they observed a 
crossover from a high--temperature regime of strong chaos to
a low--temperature weakly chaotic regime.
Specifically, above $e_c$ wave-packet excitations
of low wave-number harmonic modes were found to relax very rapidly to
a thermalized state, characterized by the equipartition of the energy
among the Fourier modes. Below $e_c$ the weakly chaotic relaxation
dynamics
exhibited a slowing-down of the energy equipartition, and the
evolution appeared increasingly far from ergodic, while decreasing the
energy density. 
In this regime, fluctuations of particles around their equilibrium
positions are quite small, so that potential (\ref{LJ}) can be 
very well approximated by its Taylor series expansion
\be
V(r) = \frac{1}{2}\omega r^2 +  
\frac{1}{3} \alpha r^3 +  
\frac{1}{4} \beta r^4 +  \cdots
\label{FPU-LJ}
\ee
where the parameters $\omega$, $\alpha$ and $\beta$ can be expressed 
in terms of $A$ and $\sigma$. This indicates that a very similar scenario
is expected to hold also in the 2d FPU model, whose potential has the form
reported on the r.h.s. of (\ref{FPU-LJ}).

More recently, the FPU model has been investigated
on a triangular 2d lattice \cite{Bti}. It has been observed that the
slowing-down of energy equipartition for sufficienty small values
of $e$ is definitely less a dramatic effect than in the square 
lattice case.
In particular, while increasing the size of the system there is
evidence that in the triangular lattice a threshold value is better
identified by the total energy $E$ , rather than by the energy density
$e$.
This implies that the effects of weak chaos should vanish in the 
thermodynamic limit much more rapidly than in the square--lattice case.
Nonetheless, all of these results can be interpreted consistently by
considering that a typical feature of lattices of anharmonic oscillators
is the appearance of long time scales for relaxation of excitations towards
a thermalized state (equilibrium) as soon as the energy is sufficiently small.
Accordingly, this fact is expected to have some important
consequences on the heat transport also in 2d lattices.
Actually, this problem has been investigated for the FPU and the Lennard-Jones
models in \cite{Lip}. Numerical simulations performed for lattice size 
$N$ up to ${\mathcal O} (10^2)$ indicate that in the strong chaotic regime 
the thermal conductivity diverges logarithmically with the
system size, $\kappa(N) \sim \ln N$. This is consistent with the theoretical
prediction of the mode--coupling theory, which is obtained by estimating the
average value of the heat--flux time correlation function at thermal
equilibrium \cite{LLP02}.
Conversely, in the weak chaotic regime the thermal conductivity seems to
diverge according to a power-law, $\kappa \sim N^{\eta}$. The
exponent $\eta$ increases while decreasing $e$ below $e_c$
and it is expected to approach unit for vanishing 
$e$, since in this limit the harmonic lattice is recovered.
In analogy with what observed in 1d, it seems reasonable
to assume that also in 2d one should eventually recover the logarithmic
divergence also in the weak chaotic regime for sufficiently large system 
size and sufficiently long
integration time. On the other hand, the practical difficulty of
performing more extended numerical simulations 
prevents the possibility of verifying this conjecture.
Moreover, it should be mentioned that there are also conflicting results 
indicating that the 2d scenario is still quite far from being fully
understood.
For instance, further numerical simulations performed
for the 2d FPU square-lattice were found to be compatible with a 
power-law divergence of the heat conductivity, with an exponent
$\alpha \approx 0.22$ \cite{GY02}. In these simulations the authors used
larger system sizes than those used in \cite{Lip}. Accordingly, their 
results should be considered more reliable for what the dependence of
$\kappa$ on $N$ is concerned, although noone among the available theoretical 
arguments supports such a prediction.
It should be also noted that in \cite{GY02} the measurement of 
$\kappa(N)$ was obtained by non-equilibrium simulations with heat baths 
at relatively moderate temperatures with respect to those employed in 
\cite{Lip}. Although the values of the temperature chosen in
\cite{GY02} are still in the range of strong chaos, they are not far
from $e_c$ and one cannot exclude that
the observed power--law behavior could be just an artifact
of finite--size corrections already effective at temperatures too close to
the so--called equipartition threshold.

\section{Conclusions}

In spite of the many efforts made in the 50 years that separate us from the
first numerical simulation performed by Fermi, Pasta and Ulam, one cannot yet
onclude that heat conduction is fully understood. We 
discussed how the simple, apparently harmless, FPU model may serve 
to illustrate how both weakly chaotic dynamics and reduced 
dimensionality affect the validity of macroscopic transport laws. 

Instances of open problems are the shape of the temperature profile which
depends on boundary conditions and coexist with energy equipartion like in
equilibrium, when temperature is constant across the system. The effective
coupling between thermal baths and  (low- and high-frequency) Fourier modes is
a related problem, whose solution might help in identifying those boundary
conditions which can minimize the contact thermal resistance. A deeper
understanding of these issues in toy models like FPU may be illuminating to
describe energy transport in ``small systems" like single--molecules or 
nanostructured materials.

The dependence of the time (and, correspondingly, the length) to reach the
asymptotic scaling regime on the energy density at low temperatures represents
another open problem. The results here reported indicate that the time needed
for a nonlinear behaviour of the hydrodynamic modes to set in increases as an
inverse power of the energy density. The scaling behaviour is compatible with
that exhibited by the times required for the chaotization of generic
trajectories, although it follows from the analysis of completely different
dynamical processes. Accordingly, we are led to conjecture that in the
thermodynamic limit the evolution is controlled by a single time scale. Last
but not least, the scenario in 2d FPU lattices is even more unclear, the nature
of the leading behaviour being still questioned.

%%%%%%%%%%%%%%%%%%%%%%%%%%%%%%%%%%%%%%%%%%%%%%%%%%%%%%%%%%%%%%%%%%

\section*{Acknowledgements}

This work is granted by the INFM-PAIS project {\it Transport phenomena in
low-dimensional structures} and is part of the PRIN2003 project {\it Order and
chaos in nonlinear estended systems} funded by MIUR-Italy. We acknowledge a
partial support from the EU network LOCNET, Contract No. HPRN-CT-1999-00163.
Part of the numerical simulation were performed at CINECA supercomputing
facility through the INFM {\it Iniziativa trasversale ``Calcolo Parallelo"}
entitled  {\it Simulating energy transport in low-dimensional systems}.


\begin{thebibliography}{00}
\bibitem{bomb} It should not be forgot that MANIAC 1 was mainly
designed for supporting researches in nuclear physics, that yielded the
production of the first atomic bomb.
\bibitem{note} A crucial, sometimes overlooked \cite{PC00}, point is that  the
microscopic expression of $J$ depends on the chosen statistical ensembles,
because of ``systematic" contributions associated with other conservation laws
that must be subtracted out \cite{G60}. For instance, expression
(\ref{jsolid}) is correct in the microcanonical ensemble with zero total
momentum, while in the canonical ensemble (for large $N$) it should be replaced 
by 
$$ 
J \; =\; {a\over 2} \sum_n ({\dot x}_{n+1} + {\dot x}_n) \, F_n
\,-\, bv_0 \langle \sum_n F_n  \rangle \quad ,
$$
with $v_0$ being the center--of--mass velocity. This is necessary to ensure 
that the autocorrelation of $J$ appearing in the Green-Kubo formula
vanishes for large times.
\end{thebibliography}
\end{document}